\documentclass[pre,twocolumn,amsmath,amssymb,superscriptaddress,showpacs,showkeys]{revtex4}
\usepackage{times}
\usepackage[dvips]{graphicx}
\usepackage[utf8]{inputenc}
\usepackage{url}

\newcommand{\pz}{p^{(0)}}
\newcommand{\rhoint}{\rho_\textrm{int}}

\begin{document}

\title{Detecting synchronization clusters in multivariate time series via coarse-graining of Markov chains}
\author{Carsten Allefeld}
\email{allefeld@igpp.de}
\affiliation{Department of Empirical and Analytical Psychophysics, Institute for Frontier Areas of Psychology and Mental Health, Wilhelmstr.~3a, 79098 Freiburg, Germany}
\author{Stephan Bialonski}
\affiliation{Department of Epileptology, Neurophysics Group, University of Bonn, Sigmund-Freud-Str.~25, 53105 Bonn, Germany}
\affiliation{Helmholtz Institute for Radiation and Nuclear Physics, University of Bonn, Nussallee 14--16, 53115 Bonn, Germany}

\begin{abstract}
Synchronization cluster analysis is an approach to the detection of underlying structures in data sets of multivariate time series, starting from a matrix $R$ of bivariate synchronization indices. A previous method utilized the eigenvectors of $R$ for cluster identification, analogous to several recent attempts at group identification using eigenvectors of the correlation matrix. All of these approaches assumed a one-to-one correspondence of dominant eigenvectors and clusters, which has however been shown to be wrong in important cases. We clarify the usefulness of eigenvalue decomposition for synchronization cluster analysis by translating the problem into the language of stochastic processes, and derive an enhanced clustering method harnessing recent insights from the coarse-graining of finite-state Markov processes. We illustrate the operation of our method using a simulated system of coupled Lorenz oscillators, and we demonstrate its superior performance over the previous approach. Finally we investigate the question of robustness of the algorithm against small sample size, which is important with regard to field applications.
\end{abstract}



\maketitle

\section{Introduction}

Studying the dynamics of complex systems is relevant in many scientific fields, from meteorology \cite{santhanam:statistics} over geophysics \cite{marwan:cross} to economics \cite{plerou:universal} and neuroscience \cite{pereda:nonlinear, zhou:hierarchical}. In many cases, this complex dynamics is to be conceived as arising through the interaction of subsystems, and it can be observed in the form of multivariate time series where measurements in different channels are taken from the different parts of the system. The degree of interaction of two subsystems can then be quantified using bivariate measures of signal interdependence \cite{brillinger:time, priestley:nonlinear, bendat:random}. A wide variety of such measures has been proposed, from the classic linear correlation coefficient over frequency-domain variants like magnitude squared coherence \cite{kay:modern} to general entropy-based measures \cite{kraskov:estimating}. A more specific model of complex dynamics that has found a large number of applications is that of a set of self-sustained oscillators whose coupling leads to a synchronization of their rhythms \cite{pikovsky:book, boccaletti:synchronization}. Especially the discovery of the phenomenon of phase synchronization \cite{rosenblum:phase} led to the widespread use of synchronization indices in time series analysis \cite{tass:detection, lachaux:measuring, mormann:mean}.

However, by applying bivariate measures to multivariate data sets an $N$-dimensional time series is described by an $N \times N$-matrix of bivariate indices, which leads to a large amount of mostly redundant information. Especially if additional parameters come into play (nonstationarity of the dynamics, external control parameters, experimental conditions) the quantity of data can be overwhelming. Then it becomes necessary to reduce the complexity of the data set in such a way as to reveal the relevant underlying structures, that is, to use genuinely multivariate analysis methods that are able to detect patterns of multichannel interaction.

One way to do so is to trace the observed pairwise correspondences back to a smaller set of direct interactions using e.g. partial coherence \cite{granger:spectral, brillinger:time}, an approach that has recently been extended to phase synchronization \cite{schelter:partial}. Another and complementary way to achieve such a reduction is cluster analysis, that is, a separation of the parts of the system into different groups, such that signal interdependencies within each group tend to be stronger than in between groups. This description of the multivariate structure in the form of clusters can eventually be enriched by the specification of a degree of participation of an element in its cluster. The straightforward way to obtain clusters by applying a threshold to the matrix entries has often been used \cite{zhou:hierarchical, kim:systematic, rodriguez:perceptions}, but it is very susceptible to random variation of the indices. As an alternative, several attempts have recently been made to identify clusters using eigenvectors of the correlation matrix \cite{plerou:random, kim:systematic, utsugi:random}, which were motivated by the application of random matrix theory to empirical correlation matrices \cite{plerou:universal, mueller:detection}.

In the context of phase synchronization analysis, a first approach to cluster analysis was based on the derivation of a specific model of the internal structure of a synchronization cluster \cite{allefeld:approach, allefeld:about}. The resulting method made the simplifying assumption of the presence of only one cluster in the given data set, and focused on quantifying the degree of involvement of single oscillators in the global dynamics. Going beyond that, the \emph{participation index} method \cite{allefeld:eigenvalue} defined a measure of oscillator involvement based on the eigenvalues and eigenvectors of the matrix of bivariate synchronization indices, and attributed oscillators to synchronization clusters based on this measure.

But despite the apparent usefulness of eigenvalue decomposition for the purposes of group identification, beyond some phenomenological evidence no good reason has been put forward why an eigenvector of a synchronization matrix should directly indicate the system elements taking part in a cluster. Moreover, in a recent survey of the performance of synchronization cluster analysis in simulation and field data \cite{bialonski:identifying} it has been shown that there are important special cases---clusters of similar strength that are slightly synchronized to each other---where the assumed one-to-one correspondence of eigenvectors and clusters is completely lost.

In this paper, we provide a better understanding of the role of eigenvectors in synchronization cluster analysis, and we present an improved method for detecting synchronization clusters, the \emph{eigenvector space} approach. The organization of the paper is as follows: In Section~II, we briefly recall the definition of the matrix of bivariate synchronization indices $R$ as the starting point of the analysis. We motivate its transformation into a stochastic matrix $P$ describing a Markov chain and detail the properties of that process. Utilizing recent results on the coarse-graining of finite-state Markov processes \cite{froyland:statistically, gaveau:dynamical, deuflhard:robust} we derive our method of synchronization cluster analysis, and we illustrate its operation using a system of coupled Lorenz oscillators. In Section~III, we compare the performance of the eigenvector space method with that of the previous approach, the participation index method \cite{allefeld:eigenvalue}, and we investigate its behavior in the case of small sample size, which is important with regard to the application to empirical data.

\section{Method}

\subsection{Measuring synchronization}

Synchronization is a generally occurring phenomenon in the natural sciences, which is defined as the dynamical adjustment of the rhythms of different oscillators \cite{pikovsky:book}. Because an oscillatory dynamics is described by a phase variable $\phi$, a measure of synchronization strength is based on the instantaneous phases $\phi_{im}$ of oscillators $i = 1 \ldots N$, where the index $m$ enumerates the values in a sample of size $n$. The nowadays commonly used bivariate index of phase synchronization strength \cite{rodriguez:perceptions, lachaux:measuring, mormann:mean, allefeld:approach, allefeld:eigenvalue, bialonski:identifying, schelter:partial} results from the application of the first empirical moment of a circular random variable \cite{mardia:directional} to the distribution of the phase difference of the two oscillators:
\begin{equation} \label{rbar}
R_{ij} = \left | \frac{1}{n} \sum_{m = 1}^n e ^ {\mathrm{i} \, (\phi_{im} - \phi_{jm})} \right |.
\end{equation}
The measure takes on values from the interval $[0, 1]$, representing the continuum from no to perfect synchronization of oscillators $i$ and $j$; the matrix $R$ is symmetric, its diagonal being composed of $1$s. Special care has to be taken in applying this definition to empirical data, because the interpretation of $R$ as a synchronization measure in the strict sense only holds if phase values were obtained from different self-sustained oscillators. 

The determination of the phase values $\phi_{im}$ generally depends on the kind of system or data to be investigated. For the analysis of scalar real-valued time series $s_i(t)$ that are characterized by a pronounced dominant frequency, the standard approach utilizes the associated complex-valued analytic signal $z_i(t)$ \cite{gabor:theory}, within which every harmonic component of $s_i(t)$ is extended to a complex harmonic. The analytic signal is commonly defined \cite{rosenblum:phase} as
\begin{equation}
z_i(t) = s_i(t) + \mathrm{i} ~ \mathrm{H} s_i(t),
\end{equation}
where $\mathrm{H} s_i$ denotes the Hilbert transform of the signal $s_i$,
\begin{equation}
\mathrm{H} s_i(t) = {1 \over \pi} \, \textrm{\scriptsize P.V.} \!\!\! \int_{-\infty}^\infty {s_i(t') \over t - t'} \mathrm{d} t',
\end{equation}
and where P.V. denotes the Cauchy principal value of the integral. The instantaneous phase of the time series is then defined as
\begin{equation}
\phi_i(t) = \arg z_i(t).
\end{equation}
Equivalently, the analytic signal can be obtained using a filter that removes negative frequency components,
\begin{equation} \label{as}
z_i(t) = \mathcal{F}^{-1} \left ( \mathcal{F} \left [ s_i(t) \right ] ~ \left [ 1 + \mathrm{sgn}(\omega) \right ] \right ),
\end{equation}
where $\mathcal{F}$ denotes the Fourier transform into the domain of frequencies $\omega$ and $\mathrm{sgn}$ denotes the sign function \cite{carmona:practical}. This definition is more useful in practice because it can be straightforwardly applied to empirical time series, which are sampled at a finite number $n$ of discrete time points $t_m$, $s_{im} = s_i(t_m)$. If several time series (realizations of the same process) are available, the obtained phase values can be combined into a single multivariate sample of phases $\phi_{im}$, where the index $m = 1 \ldots n$ now enumerates the complete available phase data.

\subsection{Cluster analysis via Markov coarse graining}

In the participation index method \cite{allefeld:eigenvalue}, the use of eigenvectors of $R$ for synchronization cluster analysis was motivated by the investigation of the spectral properties of correlation matrices in random matrix theory \cite{mueller:detection}. Another context where eigenvalue decomposition turns up naturally is the computation of matrix powers, which becomes as simple as possible using the spectral representation of the matrix.

Powers of $R$ have a well defined meaning in the special case of a binary-valued matrix ($R_{ij} \in \{0,\,1\}$), as it is obtained for instance by thresholding: The matrix entries of $R^a$ count the number of possible paths from one element to another within $a$ steps, i.e., they specify the degree to which these elements are connected via indirect links of synchrony. By analogy, we interpret $(R^a)_{ij}$ also in the general case as quantifying the degree of common entanglement of two elements $i$ and $j$ within the same web of synchronization relations. In the following we will call this quantity the $a$-step synchronization strength of two oscillators because it reduces to the original bivariate synchronization strength $R_{ij}$ in the case $a = 1$.

This synchronization strength over $a$ steps is relevant for synchronization cluster analysis, because in a system where synchronization clusters are present it is possible that the degree of direct bivariate synchrony of two elements is not very strong, but they are both entangled into the same web of links of synchrony. These indirect links, which make the two elements members of the same synchronization cluster, become visible in $R^a$. 

Moreover, with increasing power $a$ the patterns of synchrony within a cluster (the matrix columns) become more similar, approaching the form of one of the dominant eigenvectors. If there are different clusters in the system, a suitable $a$ can be chosen such that each cluster exhibits a different pattern, representing the web of synchronization relations it is comprised of. These patterns constitute \emph{signatures} by which elements can be attributed to clusters. The cluster signatures are related to the dominant eigenvectors of $R$; by transition to larger $a$ they become even more dominant, leading to an effective simplification of the matrix.

For the identification of the members of a cluster only the patterns of synchrony are relevant, while the absolute size of elements of different columns diverges with $a$, so that some sort of normalization is called for. Different normalization schemes might be used for this purpose. However, using the $L^1$-norm the procedure can be simplified, because for the normalized version of the synchronization matrix, given by 
\begin{equation} \label{tm}
P_{ij} = \frac{R_{ij}}{\sum_{i'} R_{i'j}},
\end{equation}
it holds that powers of $P$ are automatically normalized, too. Moreover, the $L^1$-normalized matrix $P$ is a column-stochastic matrix, that is, it can be interpreted as the matrix of $i \leftarrow j$ transition probabilities describing a Markov chain, whose states correspond to the elements of the original system. Via this connection, the tools of stochastic theory and especially recent work on the coarse-graining of finite-state Markov processes \cite{froyland:statistically, gaveau:dynamical, deuflhard:robust} can be utilized for the purposes of synchronization cluster analysis.

The Markov process defined in this way possesses some specific properties \cite{meyn:markov}: It is aperiodic because of the nonzero diagonal entries of the matrix, and it is in general irreducible because the values of empirical $R_{ij}$ for $i \neq j$ will also almost never be exactly zero. For a finite-state process, these two properties amount to ergodicity, which implies that any distribution over states converges to a unique invariant distribution $\pz$, corresponding to the eigenvector of $P$ for the unique largest eigenvalue $1$. This distribution can be computed from $R$ as
\begin{equation}
\pz_i = \frac{\sum_j R_{ij}}{\sum_{i'} \sum_j R_{i'j}},
\end{equation}
where the vector components of $\pz$ are denoted by $\pz_i$. With the matrix $R$ also the stationary flow given by
\begin{equation}
P_{ij} ~ \pz_j = \frac{R_{ij}}{\sum_{i'} \sum_{j'} R_{i'j'}}
\end{equation}
is symmetric, i.e., the process fulfills the condition of detailed balance $P_{ij} ~ \pz_j = P_{ji} ~ \pz_i$, which makes eigenvalues and eigenvectors of $P$ real-valued \cite{froyland:statistically}.

For the Markov process, the $a$-step synchronization strength considered above translates into transitions between states over a period of $\tau$ time steps. To compute the corresponding transition matrix $P^\tau$ the eigenvalue decomposition of $P$ is used. If $\lambda_k$ with $k = 0 \ldots N - 1$ denote the eigenvalues of $P$, and the right and left eigenvectors $p_k$ and $A_k$ are scaled such that the orthonormality relation
\begin{equation}
A_k ~ p_l = \delta_{kl}
\end{equation}
is fulfilled, the spectral representation of $P$ is given by
\begin{equation}
P = \sum_k \lambda_k ~ p_k A_k
\end{equation}
and consequently
\begin{equation}
P^\tau = \sum_k \lambda_k^\tau ~ p_k A_k.
\end{equation}
We assume that eigenvalues are sorted such that $\lambda_0 = 1 > |\lambda_1| \geq |\lambda_2| \geq \ldots \geq |\lambda_{N-1}|$. The scaling ambiguity left by the orthonormality relation is resolved by choosing
\begin{equation}
p_{ik} = \pz_i ~ A_{ki}
\end{equation}
(where $p_{ik}$ and $A_{ki}$ denote the vector components of $p_k$ and $A_k$, respectively), which leads to the normalization equations
\begin{equation} \label{ne}
\sum_i \frac{p_{ik}^2}{\pz_i} = 1
\quad \text{and} \quad
\sum_i \pz_i ~ A_{ki}^2 = 1,
\end{equation}
with the special solutions $p_{i0} = \pz_i$ and $A_{0i} = 1$. Additionally, a generalized orthonormality relation
\begin{equation} \label{go}
\sum_i \frac{p_{ik} ~ p_{il}}{\pz_i} = \delta_{kl}
\end{equation}
follows for the right eigenvectors.

The convergence of every initial distribution to the stationary distribution $\pz$ corresponds to the fact that because of non-vanishing synchronies the whole system ultimately forms one single cluster. This perspective belongs to a timescale $\tau \rightarrow \infty$, at which all eigenvalues $\lambda_k^\tau$ go to $0$ except for the largest one, $\lambda_0^\tau = 1$. In the other extreme of a timescale $\tau = 0$, $P^\tau$ becomes the identity matrix, all of its columns are different, and the system disintegrates into as many clusters as there are elements. For the purposes of cluster analysis, intermediate timescales are of interest on which many but not all of the eigenvalues are practically zero. If we want to identify $q$ clusters, we expect to find that many different cluster signatures, and that means we have to consider $P^\tau$ at a time scale where eigenvalues $\lambda_k^\tau$ may be significantly different from zero only for the range $k = 0 \ldots q - 1$.

This is achieved by determining $\tau$ such that $|\lambda_q|^\tau \approx 0$. Using a parameter $\zeta \ll 1$ chosen to represent the quantity that is considered to be practically zero (e.g. $\zeta = 0.01$), from $|\lambda_q|^\tau = \zeta$ we calculate the appropriate timescale for a clustering into $q$ clusters as
\begin{equation} \label{ts}
\tau(q) = \frac{\log \zeta}{\log |\lambda_q|}.
\end{equation}
The vanishing of the smaller eigenvalues at a given timescale describes the loss of internal differentiation of the clusters, the removal of the structural features encoded in the corresponding weaker eigenvectors. On the other hand, the differentiation of clusters from each other via the dominant eigenvectors will be the clearer the larger the remaining eigenvalues are, especially $\lambda_{q-1}^\tau$. This provides a criterion for selecting the number of clusters $q$: the clustering will be the better the larger $|\lambda_{q-1}|^{\tau(q)}$ is. Equivalently, we select $q$ based on the \emph{timescale separation factor}
\begin{equation} \label{tsf}
F(q) = \frac{\tau(q-1)}{\tau(q)}= \frac{\log |\lambda_q|}{\log |\lambda_{q-1}|},
\end{equation}
which is independent of the particular choice of $\zeta$, and invariant under rescaling of the time axis. This criterion gives a ranking of the different possible choices (from 1 to $N - 1$). The fact that $\lambda_0 = 1$ and therefore $F(1) = \infty$ implies a limitation of this approach, since the choice $q = 1$ is always characterized as absolutely optimal. Therefore the first meaningful---and usually best---choice is the second entry in the $q$ ranking list.

To determine which elements belong to the same cluster, we need a measure $d$ of the dissimilarity of cluster signatures, that is, of the column vectors of $P^\tau$. Since these vectors belong to the space of right eigenvectors of $P$, the appropriate dissimilarity metric is based on the norm corresponding to the normalization equation for right eigenvectors [Eq.\,(\ref{ne}) left]:
\begin{equation}
|| p || = \sum_i \frac{p_{ik}^2}{\pz_i}.
\end{equation}
The resulting column vector dissimilarity 
\begin{equation}
d^2(j, j') = \sum_i \frac{1}{\pz_i} \left | \left ( P^\tau \right ) _{ij} - \left ( P^\tau \right ) _{ij'} \right | ^2,
\end{equation}
has the convenient property that the dimensionality of the space within which the clustering has to be performed can be reduced, because the expression obtained by inserting the spectral representation for the matrix entries of $P^\tau$,
\begin{equation}
\left ( P^\tau \right ) _{ij} = \sum_k \lambda_k^\tau ~ p_{ik} A_{kj},
\end{equation}
simplifies to
\begin{equation}
d^2(j, j') = \sum_k |\lambda_k|^{2\tau} \left ( A_{kj} - A_{kj'} \right ) ^2
\end{equation}
using the generalized orthonormality of right eigenvectors, Eq.\,(\ref{go}). Since for appropriately chosen $\tau = \tau(q)$ contributions for larger $k$ vanish starting from $k = q$, and because $A_{0i} = 1$ for all $i$, it is sufficient to let the sum run over the range $1 \ldots q - 1$. The dissimilarity $d$ can therefore be interpreted as the Euclidean distance within a $(q - 1)$-dimensional left \emph{eigenvector space}, where each element $j$ is associated with a position vector
\begin{equation} \label{es}
\vec{o}(j) = \left ( |\lambda_k|^\tau ~ A_{kj} \right ),
\quad k = 1 \ldots q - 1.
\end{equation}

To actually perform the clustering, we can in principle use any algorithm that is designed to minimize the sum of within-cluster variances. Our implementation derives from the observation that clusters in eigenvector space form a $q$-simplex \cite{gaveau:dynamical, deuflhard:robust}. A first rough estimate of the cluster locations can therefore be obtained by searching for the extreme points of the data cloud, employing a subalgorithm described in Ref.\,\cite{deuflhard:robust}: Determine the point farthest from the center of the cloud, then the point farthest from the first one; then iteratively the point farthest from the hyperplane spanned by all the previously identified points, until $q$ points are found. Using the result of this procedure as initialization, the standard k-means algorithm \cite{macqueen:methods} that normally tends to get stuck in local minima converges in almost all cases quickly onto the correct solution.

In summary, the algorithmic steps \footnote{An implementation of the algorithm for \textsc{Matlab} and \textsc{Octave} is available from the corresponding author.} of the eigenvector space method introduced in this paper are: (1)~Calculate the matrix of bivariate synchronization indices $R_{ij}$, Eq.\,(\ref{rbar}). (2)~Convert the synchronization matrix $R$ into a transition matrix $P$, Eq.\,(\ref{tm}). (3)~Compute the eigenvalues $\lambda_k$ and left eigenvectors $A_k$ of $P$. (4)~Select the number of clusters $q$, $q > 1$, with the largest timescale separation factor $F(q)$, Eq.\,(\ref{tsf}). (5)~Determine the positions $\vec{o}(j)$, Eq.\,(\ref{es}), in eigenvector space for $\tau = \tau(q)$, Eq.\,(\ref{ts}). (6)~Search for $q$ extreme points of the data cloud. (7)~Use these as initialization for k-means clustering.

\subsection{Illustration of the eigenvector space method}

\begin{figure}
\includegraphics{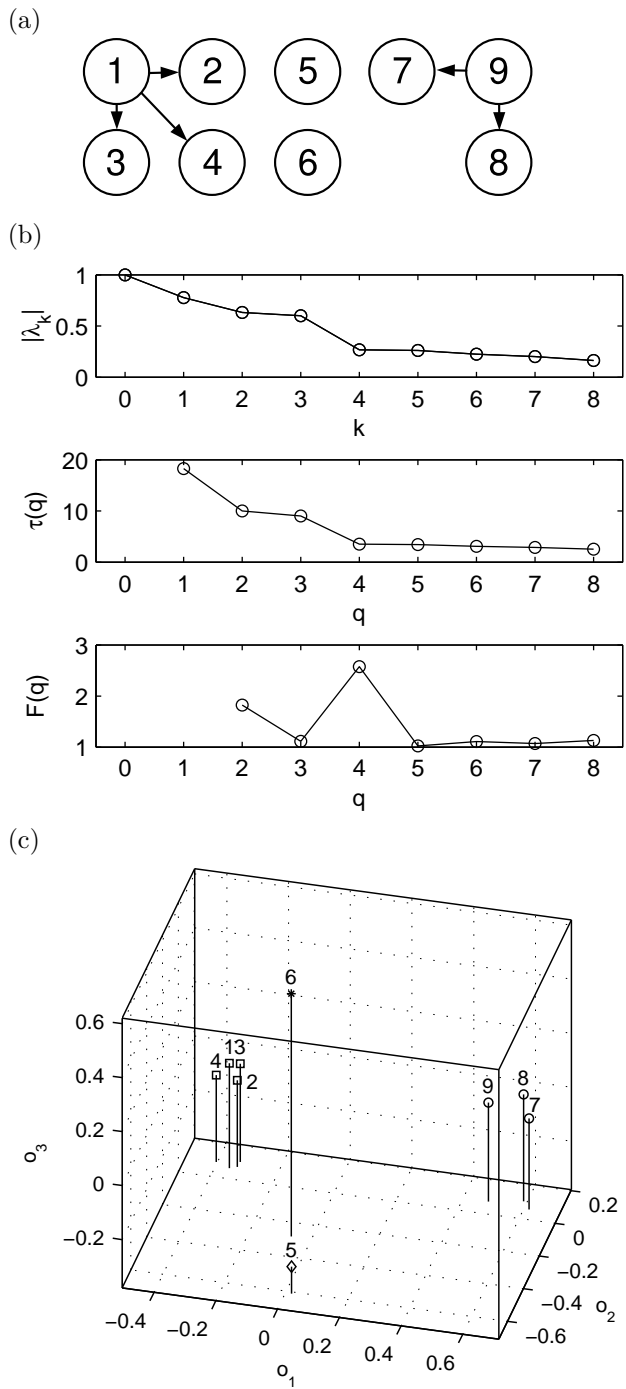}
\caption{Application of the eigenvector space method to a system of nine partially coupled Lorenz oscillators. (a)~Coupling configuration: The left group of four oscillators is driven by \#1, the right group of three driven by \#9, the remaining two are uncoupled. (b)~Eigenvalues $\lambda_k$, timescales $\tau(q)$, and timescale separation factors $F(q)$. The maximal separation factor $F(4)$ indicates the presence of four clusters. (c)~Positions attributed to oscillators in 3-dimensional eigenvector space $(o_1, o_2, o_3)$. The clustering by the k-means algorithm results in a cluster composed of oscillators \#1--4 ($\scriptstyle\square$), two single-element clusters consisting in oscillators \#5 ($\diamond$) and \#6 ($\ast$), respectively, and a cluster composed of oscillators \#7--9 ($\circ$).}
\label{ex}
\end{figure}

In order to illustrate the operation of the method, we apply it to multivariate time series data obtained from simulated nonlinear oscillators, coupled in such a way as to be able to observe synchronization clusters of different size as well as unsynchronized elements. The system consists of $N=9$ Lorenz oscillators that are coupled diffusively via their $z$-components:
\begin{eqnarray}
\dot{x}_j &=& 10\,(y_j - x_j), \nonumber \\
\dot{y}_j &=& 28\,x_j - y_j - x_j z_j,\\
\dot{z}_j &=& -\tfrac{8}{3}\,z_j + x_j y_j \qquad + ~ \epsilon_{ij} \, (z_i - z_j). \nonumber
\end{eqnarray}
The coupling coefficients $\epsilon_{ij}$ were chosen from $\{0, 1\}$ to implement the coupling configuration depicted in Fig.\,\ref{ex}\,(a), such that oscillators \#2--4 are unidirectionally driven by \#1, oscillators \#7 and 8 driven by \#9, and \#5 and 6 are uncoupled. These differential equations were numerically integrated using a step size of $\Delta t = 0.01$, starting from randomly chosen initial conditions. After discarding an initial transient of $10^4$ data points, further $4 \times 10^4$ samples entered data processing. Instantaneous phases $\phi_{jm}$ of oscillators $j$ at time points $t_m = (m - 1) \, \Delta t$ were determined from the $z$-components using the analytic signal approach after removal of the temporal mean, and bivariate synchronization strengths $R_{ij}$ were computed.

The outcomes of the eigenvector space method applied to the resultant matrix of bivariate indices are presented in Fig.\,\ref{ex}. Figure~\ref{ex}\,(b) shows the spectrum of eigenvalues $\lambda_k$ of the transition matrix $P$ with the corresponding timescales $\tau(q)$ and timescale separation factors $F(q)$. A gap in the eigenvalue spectrum between indices $k = 3$ and $4$ translates into a maximum timescale separation factor for $q = 4$, which recommends a search for four clusters in the eigenvector space for timescale $\tau = 3.5$. This 3-dimensional space is depicted in Fig.\,\ref{ex}\,(c), where the expected grouping into four clusters can be clearly recognized in the arrangement of elements $j$ with positions $\vec{o}(j)$. These four clusters that correspond to the two groups of driven oscillators and the two uncoupled oscillators (each of which forms a single-element cluster) are easily identified by the k-means algorithm. The results shown here were obtained using $\zeta = 0.01$; alternative choices of $0.1$ and $0.001$ yielded the same clustering.

\section{Performance}

To assess the performance of the eigenvector space method introduced in this paper, we compare it with the previous approach to synchronization cluster analysis based on spectral decomposition. For reference, we briefly recollect the important details.

The participation index method \cite{allefeld:eigenvalue} is based on the eigenvalue decomposition of the symmetric synchronization matrix $R$ itself, into eigenvalues $\eta_k$ and $L^2$-normalized eigenvectors $v_k$. Each of the eigenvectors that belong to an eigenvalue $\eta_k > 1$ is identified with a cluster, and a system element $j$ is attributed to that cluster $k$ in which it participates most strongly, as determined via the participation index
\begin{equation}
\Pi_{jk} = \eta_k \, v_{jk}^2,
\end{equation}
where $v_{jk}$ are the eigenvector components of $v_k$. The method performs quite well in many configurations, but it encounters problems when confronted with clusters of similar strength that are slightly synchronized to each other, which was demonstrated in Ref.\,\cite{bialonski:identifying} using a simulation.

\begin{figure}
\includegraphics{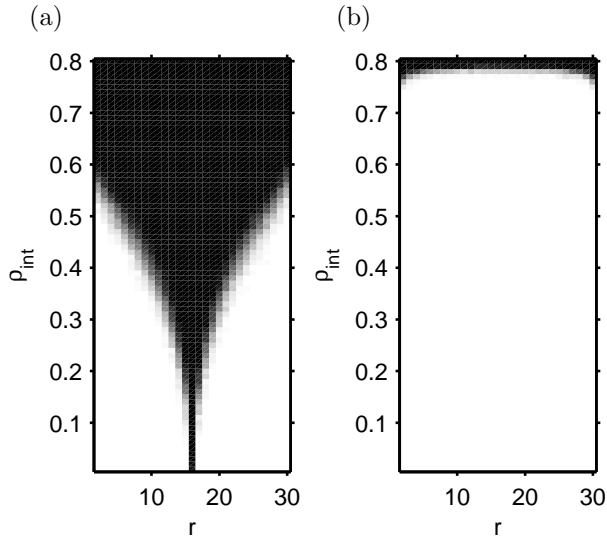}
\caption{Comparative performance of the participation index~(a) and the eigenvector space method~(b). The methods are tested on a system of $N=32$ elements, divided into two clusters containing $r$ and $N-r$ elements, respectively. The inter-cluster synchronization strength $\rhoint$ is varied from $0$ up to the value of intra-cluster synchronization $0.8$. Synchronization matrices are generated based on samples of size $n = 200$. The quantity shown is the relative frequency (over 100 trials) with which the respective algorithm failed to recover exactly the given two-cluster structure; it is visualized in gray scale, covering the range from 0 (white) to 1 (black). Comparison shows that in a large area along $r=16$ where the participation index method fails, the eigenvector space method introduced in this paper performs perfectly.}
\label{tc}
\end{figure}

Here we employ a refined version of that simulation to compare the two methods. We consider a system of $N = 32$ oscillators forming two clusters, and check whether the methods are able to detect this structure from the bivariate synchronization matrix $R$ for different degrees of inter-cluster synchrony $\rhoint$. The cluster sizes are controlled via a parameter $r$, such that the first cluster comprises elements $j = 1 \ldots r$, the second $(r+1) \ldots N$.

To be able to time-efficiently perform a large number of simulation runs and to have precise control over the structure of the generated synchronization matrices, we do not implement the system via a set of differential equations. Instead, our model is parametrized in terms of the population value of the bivariate synchronization index, Eq.\,(\ref{rbar}),
\begin{equation} \label{rho}
\rho_{ij} = \left | \left \langle \exp \left [ \mathrm{i} \, (\phi_{i} - \phi_{j}) \right ] \right \rangle \right |
\end{equation}
(where $\langle \cdot \rangle$ denotes the expectation value), which is the first theoretical moment of the circular phase difference distribution \cite{mardia:directional}. For $i, j$ within the same cluster $\rho_{ij}$ is fixed at a value of $\rho_1 = \rho_2 = 0.8$. For inter-cluster synchronization relations it is set to $\rhoint$, which is varied from $0$ up to $0.8$ such that the two-cluster structure almost vanishes.

To be able to properly account for the effect of random variations of $R_{ij}$ around $\rho_{ij}$ due to finite sample size $n$ we generated samples of phase values, using an extension of the single-cluster model introduced in Ref.\,\cite{allefeld:approach}: The common behavior of oscillators within each of the two clusters is described by cluster phases $\Phi_1$ and $\Phi_2$. The phase differences between the members of each cluster and the respective cluster phase,
\begin{equation}
\Delta \phi_{j} = \left \{
\begin{array}{ll}
\phi_j - \Phi_1  &  \text{ for } j = 1 \ldots r,  \\
\phi_j - \Phi_2  &  \text{ for } j = (r + 1) \ldots N,
\end{array}
\right .
\end{equation}
as well as the phase difference of the two cluster phases,
\begin{equation}
\Delta \Phi = \Phi_2 - \Phi_1,
\end{equation}
are assumed to be mutually independent random variables, distributed according to wrapped normal distributions \cite{mardia:directional} with circular moments $\rho_\mathrm{1C}$,  $\rho_\mathrm{2C}$, and  $\rho_\mathrm{CC}$, respectively. Since the summation of independent circular random variables results in the multiplication of their first moments \cite{allefeld:approach}, for the relation of model parameters and distribution moments holds
\begin{eqnarray}
& \rho_1 = \rho_\mathrm{1C}^2, &  \nonumber \\
& \rho_2 = \rho_\mathrm{2C}^2, &  \\
& \rhoint = \rho_\mathrm{1C} ~ \rho_\mathrm{CC} ~ \rho_\mathrm{2C}. &  \nonumber
\end{eqnarray}
For the performance comparison of the two methods, $n = 200$ realizations of this model of the multivariate distribution of phases $\phi_j$ were generated for each setting of the parameters, and synchronization indices $R_{ij}$ were calculated via Eq.\,(\ref{rbar}).

The clustering results are presented in Fig.\,\ref{tc} for the participation index and the eigenvector space method (using $\zeta = 0.01$). The quantity shown is the relative frequency (over 100 instances of the matrix $R$) of the failure to identify correctly the two clusters built into the model. Figure~\ref{tc}\,(a) shows that the participation index method fails systematically within a region located symmetrically around $r = N/2 = 16$ (clusters of equal size). The region becomes wider for increasing $\rhoint$ but is already present for very small values of the inter-cluster synchronization strength. In contrast, the eigenvector space approach (b) is able to perfectly reconstruct the two clusters for all values of $r$ up to very strong inter-cluster synchronization. It fails to correctly recover the structure underlying the simulation data only in that region where inter-cluster synchronization indices attain values comparable to those within clusters, i.e., only where there are no longer two different clusters actually present. These results demonstrate that the eigenvector space method is a clear improvement over the previous approach.

\begin{figure}
\includegraphics{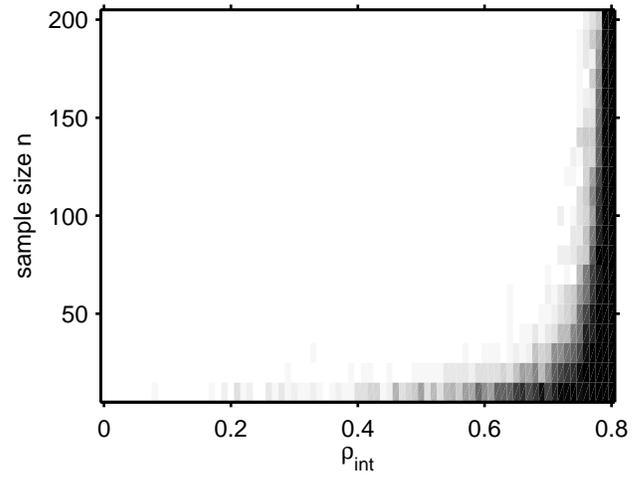}
\caption{Performance of the eigenvector space method depending on the sample size $n$, investigated using the two-cluster system of Fig.\,\ref{tc} for different values of the inter-cluster synchronization strength $\rhoint$. The quantity shown here is the proportion of values of the parameter $r$ (controlling cluster sizes) at which the two-cluster structure failed to be recovered, visualized on a gray scale from 0 (white) to 1 (black). The plot shows that the ability of the eigenvector space method to correctly identify clusters up to high values of $\rhoint$ breaks down only for very small sample sizes $n$.}
\label{ss}
\end{figure}

For real-world applications, a time series analysis method has to be able to work with a limited amount of data. In the case of synchronization cluster analysis, a small sample size attenuates the observed contrast between synchronization relations of different strength, making it harder to discern clusters. Using the two-cluster model described above, in a further simulation we investigated the effect of the sample size $n$ on the performance of the eigenvector space method. Parameters $\rhoint$ and $r$ were varied as before, and synchronization matrices were generated for $n = 10 \ldots 200$ (in steps of 10). For each value of $\rhoint$ and $n$, as a test quantity the proportion of $r$-values for which the algorithm did not correctly identify the two clusters was calculated. The result shown in Fig.\,\ref{ss} demonstrates that the performance of the eigenvector space method degrades only very slowly with decreasing sample size. The method seems to be able to provide a meaningful clustering down to a data volume of about $n = 30$ independent samples, making it quite robust against small sample size.

\section{Conclusion}

We introduced a method for the identification of clusters of synchronized oscillators from multivariate time series. By translating the matrix of bivariate synchronization indices $R$ into a stochastic matrix $P$ describing a finite-state Markov process, we were able to utilize recent work on the coarse-graining of Markov chains via the eigenvalue decomposition of $P$. Our method estimates the number of clusters present in the data based on the spectrum of eigenvalues, and it represents the synchronization relations of oscillators by assigning to them positions in a low-dimensional space of eigenvectors, thereby facilitating the identification of synchronization clusters. We showed that our approach does not suffer from the systematic errors made by a previous approach to synchronization cluster analysis based on eigenvalue decomposition, the participation index method. Finally, we demonstrated that the eigenvector space method is able to correctly identify clusters even given only a small amount of data. This robustness against small sample size makes it a promising candidate for field applications, where data availability is often an issue. Concluding we want to remark that though the method was described and assessed in this paper within the context of phase synchronization analysis, our approach might also give useful results when applied to other bivariate measures of signal interdependence.

\begin{acknowledgments}
The authors would like to thank Harald Atmanspacher, Peter beim Graben, Mar{\'i}a Herrojo Ruiz, Klaus Lehnertz, Christian Rummel, and Ji{\v r}{\'i} Wackermann for comments and discussions.
\end{acknowledgments}

\end{document}